\pdfoutput=1
\documentclass[
twocolumn,
amsmath,
amssymb,
aps,
prb
]{revtex4-1}

\usepackage{subfigure}
\usepackage[pdftex]{graphicx,color}

\begin{document}

\title{Nanowire terahertz quantum cascade lasers}

\author{Thomas Grange}
\email{thomas.grange@neel.cnrs.fr}
\altaffiliation{Present address: Institut N\'{e}el--CNRS, 25 av. des Martyrs, 38042 Grenoble, France.}
\affiliation{
 Walter Schottky Institut, Technische Universit\"{a}t M\"{u}nchen, Am Coulombwall 4, D-85748, Garching, Germany
}

\date{\today}

\begin{abstract}

Quantum cascade lasers made of nanowire axial heterostructures
are proposed.
The dissipative quantum dynamics of their carriers 
is theoretically investigated using non-equilibrium Green functions.
Their transport and gain properties are calculated for varying nanowire thickness, from the classical-wire regime to the quantum-wire regime.
Our calculation
shows
that the
lateral quantum confinement provided by the nanowires
allows an increase of the maximum operation temperature and
a strong reduction of the current density threshold compared to conventional terahertz quantum cascade lasers.

\end{abstract}

\maketitle

Quantum cascade lasers (QCLs) \cite{Faist94} are semiconductor unipolar devices emitting coherent infrared radiation.
Their active regions consist of stacked planar quantum wells (QWs) forming a superlattice along the growth direction. At first sight, a large part of the physical properties of QCLs can be qualitatively understood in a one-dimensional picture, in particular for quantum confinement, tunneling processes and radiative transitions. Nevertheless, the free in-plane motion of the carriers provides a continuous energy reservoir (i.e. subbands) which plays a major role in the energy and phase relaxation processes.
For QCLs operating in the terahertz (THz) range \cite{Kohler02,williams07},
due to the relatively small energy separation between the subbands associated with the lasing levels, the non-radiative scattering processes are very efficient and are responsible for intrinsic limitations
in terms of quantum efficiency and maximum operation temperature.

Reducing the dimensionality
of electronic systems is
generally a
way to
increase the carrier lifetimes \cite{bockelmann90, sauvage02,pandey2008slow, zibik09}.
A strong  magnetic field applied along the growth axis of a QCL quantizes the lateral motion in terms of Landau levels.
Their energy tuning with magnetic field has been shown to be responsible for modulation of the level lifetimes \cite{becker2002gaas,becker2004electron}, allowing to demonstrate lasing up to higher temperature (225~K in ref.~\cite{wade2008magnetic}) than in usual THz QCLs (199~K in ref.~\cite{fathololoumi2012terahertz}). However, Landau levels remain highly degenerate
so that
their ideal discrete density of states is strongly broadened by disorder effects \cite{leuliet2006electron}.
On the other hand, QCLs based on 0D nanostructures such as self-assembled or lithographically defined quantum dots (QDs) have been proposed \cite{Wingreen97,hsu2000intersubband,Dmitriev05,vukmirovic2008electron} in order to achieve low threshold current and high operation temperature.
However,
a direct comparison of the device performances with respect to conventional THz QCLs is lacking, and
the intermediate regime between the limiting cases coupled 0D QDs and coupled 2D QWs remains to be investigated.

In this letter we propose QCLs based on nanowire (NW) superlattice heterostructures.
Using the formalism of non-equilibrium Green functions (NEGF),
we calculate charge transport and THz gain
in
laterally-quantized quantum cascade (QC) structures.
We investigate the dynamics of charge carriers in QC structures having lateral dimensionality ranging from 2D to 0D, providing a direct comparison of NW QCLs and conventional QCLs.
Our calculations predict that
below a critical thickness
NW THz QCLs can achieve
higher gain, higher operation temperature and lower current densities
than conventional THz QCLs.
For thin nanowires, large tunneling barriers in the axial heterostructure are shown to be necessary to avoid electrical instability owing to the increase of coherence lengths.

\begin{figure}
\begin{centering}
\includegraphics[]{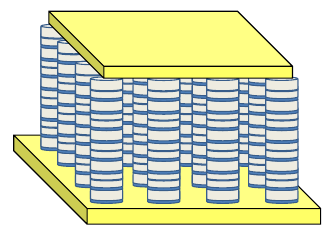}
\end{centering}
\caption{Schematic of what a nanowire terahertz quantum cascade laser could look like. The active regions consists of nanowires containing axial periodic heterostructures. The space between the nanowires could be filled with a low-loss polymers \cite{krall2014subwavelength}. The nanowire array could be sandwiched in a metal-metal waveguide.}
\label{schematic}
\end{figure}

A schematic of the proposed NW-QCL is presented on Fig~\ref{schematic}. The active region consists of an array of nanowires containing axial superlattice heterostructures.
Experimentally, a micropillar version of such THz QCL geometry has been recently reported in \cite{krall2014subwavelength}.
We have recently studied theoretically electron transport in nanowire superlattices where one period consists of a single well and a single barrier \cite{grange2014SLs}. The charge transport along the superlattice has been shown to depend dramatically on the nanowire lateral dimensions, with increasing coherence lengths predicted for decrasing nanowire thickness.
The underlying reason is that lateral energy quantization greatly limits the scattering mechanisms allowed by energy conservation, resulting in longer lifetimes and coherence lengths.

We consider one of the nanowire of Fig~\ref{schematic}. The electronic structure is modeled within the effective mass approximation. The NW surface is assumed to be perfectly passivated so that no charge depletion or accumulation effects occur.
We consider an electron wave function basis set of the form $\Psi_{\alpha n} = \psi_{\alpha}(z)\phi_{n}(\rho,\theta)$, where $\phi_{n}$ are the solutions of the Schr\"{o}dinger equation on a homogeneous disc (ie the assumed NW cross-section) with eigenvalues $E_n$ and the $\psi_{\alpha}$ form a periodic basis set of the low energy states along the $z$ superlattice axis.
The electron dynamics is calculated within the NEGF formalism in which the  perturbative scattering processes are treated within the self-consistent Born approximation (SCBA).
Self-consistent (SC) NEGF implementation of dissipative quantum transport in semiconducting heterostructures
\cite{lake97, lee02, Vukmirovic07, kubis09, grange2014SLs} 
allows  a coherent
and non-Markovian treatment of  the charge dynamics,
where the scattering self-energies can be conveniently treated on a microscopic basis. 
The SC-NEGF formalism has the advantage to correctly describe the intrinsic relation between scattering processes and line broadening effects, which is crucial here as we are interested in dimensionality transition and gain broadening.

We take into account the scattering mechanisms that are known to be important in QW QCLs heterostructures: longitudinal optical (LO) and acoustical phonon emission/absorption, interface roughness, charged impurities and alloy disorder. Electron-electron scattering is not considered. It is expected to play a minor role in the resonant-phonon Thz QCL design studied below \cite{jirauschek2009monte}.
In addition,
scattering mechanisms specific to NWs and/or quasi 0D systems are also considered.
The NW surface roughness is taken into account as an additional elastic scattering mechanism. NW phonon confinement effects are considered and we calculate the coupling of electrons to longitudinal optical (LO) and
surface optical (SO) phonons \cite{xie2000bound}, as well as confined acoustic phonons \cite{grange2014SLs}.

In quasi 0D systems,
early studies have predicted a phonon bottleneck as soon as discrete electronic transitions are detuned from the optical or acoustical phonon frequencies \cite{bockelmann90,Benisty91}.
More recent studies have demonstrated that 
(i) direct LO-phonon emission/absorption processes are totally suppressed even in case of electron--LO-phonon resonances
due to the strong coupling regime between electrons and LO-phonons \cite{hameau99}, and that (ii) the dominant energy relaxation of terahertz excitations is the polaron anharmonicity, i.e. an indirect scattering process involving electron--LO-phonon polar interaction as well as anharmonic couplings of LO-phonons with other phonon modes \cite{sauvage02, grange2007polaron, zibik09}.
This mechanism is included here, beyond a phenomenological broadening of the optical phonon GFs
\cite{Vukmirovic07},
by calculating the phononic self-energies of the optical phonons  due to their anharmonic coupling with the other phonon modes \cite{grange2014SLs}.

The elastic scattering mechanisms induced by disorder effects (here interface roughness, surface roughness, alloy disorder and charged impurities) are usually treated within the SCBA in planar QW heterostructures \cite{lake97,lee02, kubis09}.
However, the SCBA only accounts for an incoherent treatment of the disorder effects.
Here, as the NW gets thinner,
the elastic scattering becomes mainly coherent,
as the disorder potential fluctuations
become larger than the phonon-limited broadening.
In the present work,
a coherent treatment of the disorder potential is made as reported in Ref.~\cite{grange2014SLs}.

\begin{figure}
\begin{centering}

\includegraphics[]{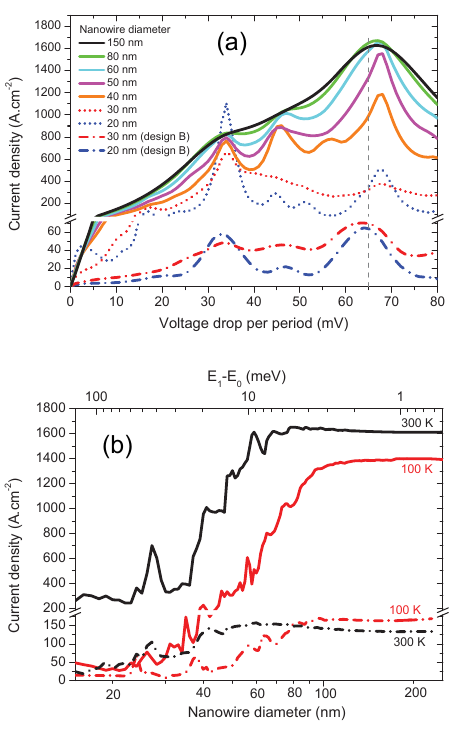}
\label{Fig2}
\end{centering}

\caption{Transport properties of NW QCLs. Axial designs A, B$_{300}$ and B$_{100}$ are studied with layer thickness of respectively 14/\textbf{1.5}/11.2/\textbf{3.5}/\underline{22}/\textbf{3.5}, 14.4/\textbf{1.5}/11.5/\textbf{7}/\underline{23}/\textbf{7} and 14.5/\textbf{1.5}/11.6/\textbf{7}/\underline{22.5}/\textbf{7} (barriers are in bold fonts and doped layer underlined). (a) Current density as a function of the voltage drop per period for various NW diameters at lattice temperature of 300 K, for the design A (solid and dotted lines for respectively stable and unstable operating regions) and for the design B$_{300}$ (dashed-dotted lines). The designed operating bias (65 mV) is indicated by a vertical dotted line. (b) Current density as a function of the NW diameter for fixed operating bias voltage at lattice temperatures of 100 K (red lines) and 300 K (black lines) for the design A in solid lines and the designs B in dashed-dotted lines (B$_{100}$ at 100K and B$_{300}$ at 300K). The top axis indicates the energy separation between the NW lateral ground state $\phi_0$ and the first excited state $\phi_1$.
}
\label{transport}
\end{figure}

We consider here the In$_{0.53}$Ga$_{0.47}$As/GaAs$_{0.51}$Sb$_{0.49}$ material system lattice matched to InP,
in which a conventional THz QCL operating up to 142~K has recently been demonstrated \cite{deutsch:211117}.
This system could be used to realize NW QCLs in a top-down approach \cite{krall2014subwavelength}, while bottom-up approaches might be envisaged on the longer term with similar material systems \cite{samuelson2004semiconductor,krogstrup2009junctions}.
We consider below
3-well resonant phonon designs (A and B),
similar to the one holding operation temperature records in conventional THz QCLs \cite{fathololoumi2012terahertz}.
The layer sequence in nanometers of the design A
is 14/\textbf{1.5}/11.2/\textbf{3.5}/\underline{22}/\textbf{3.5} with barriers indicated in bold fonts and the doped layer underlined. The equivalent areal doping density per period is set to $3\times 10^{10}$ cm$^{-2}$. The interface and surface roughnesses are both assumed to have
a rms
fluctuating amplitude of 3~\AA{} with correlation lengths of 8~nm.
The filling factor of the NW array is assumed to be unity for simplicity in the treatment of Coulombic interactions and for clear comparison with the conventional QCL geometry.
All the remaining parameters in our model are material-dependent and we take standard values from the literature \cite{vurgaftman2001band,levinshtein1996handbook}.

Fig.~\ref{transport} shows the calculated transport properties of the NW QCLs. 
The current--voltage characteristics calculated for various NW diameters at room temperature are shown on Fig.~\ref{Fig2}a. Fig.~\ref{Fig2}b shows the current density as a function of the NW diameter for a fixed applied bias
of 65~mV per period for various temperatures.
For sufficiently thick wires the calculated current densities saturate on Fig.~\ref{Fig2}b and all the IV curves for diameters larger than $\sim$150 nm (not shown on Fig.~\ref{Fig2}a) overlap. 
This corresponds to the 2D regime of the lateral motion, i.e. when the lateral quantization effects are negligible.
When the NW diameter is reduced below a temperature-dependent critical diameter, variations of the current density are observed on Fig.~\ref{Fig2}b corresponding to the appearance of lateral dimensional effects.
In this quantum wire regime,
the I--V characteristic possesses more distinct peaks
as the reduction of the scattering processes allows adiabatic transport over larger distances.
In fact, in the 2D regime the lateral motion provides an energy reservoir towards which the excess axial energy is easily transfered via the disorder potentials. As the lateral motion gets quantized, this elastic energy exchange tends to be suppressed \cite{grange2014SLs}. As the remaining scattering processes are mainly due to optical phonons, a strong current peak remains for potential drop per period matching the optical phonon energy (34 meV), whereas the current is strongly reduced for other bias voltage.
For thin diameters, as this parasitic current peak at 34 mV becomes larger than the current at the laser operating bias (around 65 meV), electrical instability can occur in the design A.
In order to restore electrical stability, designs B differing from the design A mainly by twice thicker injector and extractor tunneling barriers (7 nm instead of 3.5 nm) are investigated. Owing to the sharper tunneling resonances and the effective mass temperature dependence, two slighty different designs B$_{300}$ (B$_{100}$) with layer sequence in nm 14.4/\textbf{1.5}/11.5/\textbf{7}/\underline{23}/\textbf{7} (14.5/\textbf{1.5}/11.6/\textbf{7}/\underline{22.5}/\textbf{7}) are studied at respectively 300 K and 100 K.
As shown on Fig.~\ref{Fig2}a at 300 K, the weaker tunneling couplings in designs B allow a reduction of the current leakage peak at the optical phonon energy.
A further insight into the nature of transport is provided by the analysis of the spectral function (not shown here).
For the thick injector and collector barriers in the design A, anticrossings in the spectral function at the tunneling resonances are observed only for NW diameters smaller than $\sim35$ nm ($\sim60$ nm) at 300 K (100K respectively), indicating than the nature of these resonant tunneling processes evolves from an incoherent to a coherent regime as the NW thickness is reduced.

\begin{figure}
\begin{centering}
\includegraphics[]{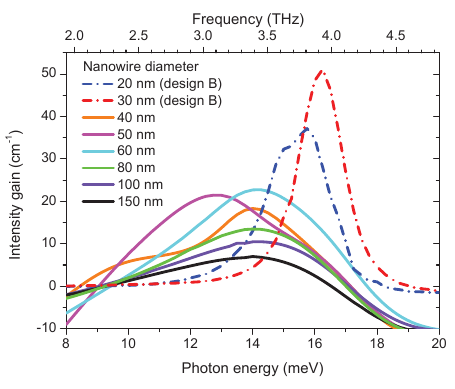}
\par\end{centering}
\caption{Intensity gain spectra at 300~K calculated for various NW diameter for the design A (solid lines) and the design B$_{300}$ (dashed-dotted lines).}
\label{gain_curves}
\end{figure}

\begin{figure}
\begin{centering}
\includegraphics[]{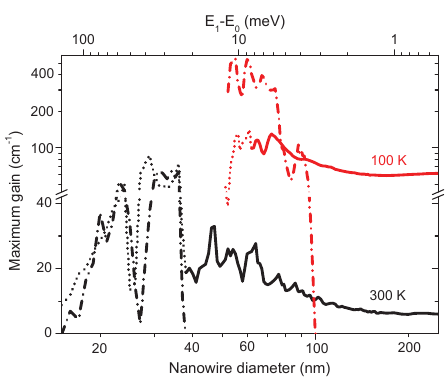}
\par\end{centering}
\caption{Maximum intensity gain as a function of the NW diameter at 300K (black lines) and 100K (red lines) for the design A (solid and dotted lines for respectively electrical stable and unstable gain regions) and for the designs B (dashed-dotted lines, only for NW diameters satisfying electrically stability).}
\label{Max_gain}
\end{figure}

We calculate the THz gain in linear response with a self-consistent treatment of the induced self-energies (i.e. including vertex corrections) \cite{wacker2002gain}, which has been shown to be necessary to correctly treat the correlation in the scattering potentials \cite{banit2005self}.
The calculated intensity gain spectra at the optimal operating bias are shown on Fig.~\ref{gain_curves} at 300~K. The peak values of the gain
are plotted on Fig.~\ref{Max_gain} as a function of the NW diameter at 100~K and 300~K.
The gain of the design A in the QW limit reaches 6 cm$^{-1}$ at room temperature, which is below the typical reported cavity losses \cite{burghoff:261111}. As the NW diameter decreases into the quantum wire regime, the gain tends to increase. This increase is attributed to the overall decrease of scattering mechanism efficiency with decreasing dimensionality, and hence increase in the lifetimes and coherence times.
Peaks in the gain are observed for certain NW diameters on Fig.~\ref{Max_gain}. This is due to the interplay between the energy of the axial heterostructure and the lateral energy quantization, the former being tuned by the NW diameter while the first is fixed.
A narrowing of the gain peaks with decreasing NW diameter is also observed on Fig.~\ref{gain_curves}. This is attributed to a decrease of the scattering process efficiency and consequently of the homogeneous transition linewiths.
Inhomogeneous broadening effects are found to remain smaller than homogeneous ones for NW diameters down to ~20 nm.
Fig.~\ref{Max_gain} allows to predict the required NW diameters in order to obtain lasing. Assuming cavity losses equivalent to an absorption of 15 (30) cm$^{-1}$, it shows that THz lasing at 300~K requires NW diameters smaller than 70 (38)~nm.
As the NW diameter is further reduced below 38~nm at 300~K (62~nm at 100~K), 
the design A becomes electrically unstable (dotted lines on Fig.~\ref{Max_gain}).
The design B is stable down to thinner NW diameters and its gain for NW diameter between 30 and 36 nm reaches values around 60 cm$^{-1}$ at room temperature, which is equivalent to the calculated value at 100~K in the 2D limit.
For NW diameters below 20~nm,
inhomogeneous energy fluctuations start to dominate over homogeneous broadening
, so that the gain strongly decreases due to the large energy fluctuations and the consequent misalignment of the QCL levels.
 The calculated gain for the designs B (B$_{300}$ and B$_{100}$) is found to be more sensitive to the NW diameter with respect to design A: it is positive below a critical diameter (38~nm at 300~K and 100~nm at 100~K), and negative beyond up to the 2D limit. Indeed,
coherent propagation through their large injector and collector tunneling barriers occurs only in the quantum wire regime, whereas in the 2D limit the  tunneling is mainly incoherent and not efficient enough to allow population inversion. A further advantage of the designs B$_{300}$ and B$_{100}$ in the quantum wire regime resides in their much smaller current densities threshold (more than one order of magnitude smaller) than the design A operating in the 2D limit.

In summary, we have predicted the transport properties of NW QC structures as well as their THz response.
We have shown that NWs QC structures allow higher THz gain and together with lower current densities compared to conventional planar QCs structures.
Large tunneling barriers are shown to be necessary in the thin NW regime in order to maintain electrical stability as coherent propagation lengths are increased.
Finally, the present work provides the geometrical requirements of NW QC structures in order to achieve THz stimulated emission at room temperature.

This work has been supported by the Alexander von Humboldt foundation and the Austrian Science Fund FWF (SFB IR-ON). S. Birner, M. Branstetter, C. Deutsch, P. Greck, G. Koblm\"{u}ller, M. Krall, T. Kubis, S. Rotter, K. Unterrainer and P. Vogl are gratefully acknowledged for fruitful discussions.

\bibliographystyle{apsrev}

\end{document}